# A Comparative Study on $q$-Deformed Fermion Oscillators

**Abdullah Algin**

Department of Physics, Eskisehir Osmangazi University,
Meselik, 26480-Eskisehir, Turkey
e-mail: aalgin@ogu.edu.tr

**Abstract**

In this paper, the algebras, representations, and thermostatistics of four types of fermionic $q$-oscillator models, called fermionic Newton (FN), Chaichian-Kulish-Ng (CKN), Parthasarathy-Viswanathan-Chaichian (PVC), Viswanathan-Parthasarathy-Jagannathan-Chaichian (VPJC), are discussed. Similarities and differences among the properties of these models are revealed. Particular emphasis is given to the VPJC-oscillators model so that its Fock space representation is analyzed in detail. Possible physical applications of these models are concisely pointed out.

# 1 Introduction

After the pioneering studies of Biedenharn [1] and Macfarlane [2] on the one-dimensional bosonic $q$-oscillators, a parallel research direction has been naturally come out as fermionic $q$-oscillators [3-28]. In the meantime, statistical and thermodynamical properties of such fermionic $q$-oscillator systems have been extensively investigated in the literature [29-38]. Such developments have also led to the discussion of possible generalizations of classical symmetry properties of the system under consideration. The most commonly studied model in this framework is the $SU_q(d)$-covariant one- or two-parameter deformed bosonic and fermionic oscillator algebras. For instance, it was shown in [39-44] that the high- and low-temperature thermostatistical behaviors of the one- and two-parameter deformed fermion gas models changed radically by invoking the quantum group symmetry to the system under consideration.

Furthermore, some of the earlier applications of fermionic quantum algebras should be mentioned: For instance, they were used to discuss some hadronic properties such as the dynamical mass generated for quarks and pure nuclear pairing force version of the Bardeen-Cooper-Schrieffer (BCS) many-body formalism [45-48]. They were also used to understand higher-order effects in the many-body interactions in nuclei [49,50]. By means of a $q$-fermionic algebra, one could also describe interactions between fermions and bosons. Thus, the present study on $q$-deformed fermionic oscillator algebra models may be used to approximate some complex behaviors in similar physical systems, and it will serve to give the importance of $q$-deformations of fermions to some extent.

In this work, we aim to compare quantum algebraic and thermostatistical properties of systems arising from the four types of the $q$-deformed fermionic oscillator algebra models, called fermionic Newton (FN), Chaichian-Kulish-Ng (CKN), Parthasarathy-Viswanathan-Chaichian (PVC), Viswanathan-Parthasarathy-Jagannathan-Chaichian (VPJC). In this framework, we should emphasize that the first three of these models are mostly studied in the literature. Although, it has different and interesting properties, the VPJC-oscillators model is relatively less studied and is not so popular among the others. Due to this fact, we wish to focus on this model in the last part of this work. Historically, this model was first introduced by Viswanathan *et al*. [7] and also, its some of the statistical properties was further investigated by Chaichian *et al*. [8] Therefore, we call them as the VPJC-oscillators model.

The paper is organized as follows: In section 2, we discuss the properties of  the FN-, CKN-, PVC-, VPJC-oscillator models. In particular, we investigate the VPJC-oscillator model so that its Fock space representation is analyzed in detail. In section 3, we reveal similarities and differences among the properties of these models. Also, possible physical applications of these models are concisely pointed out. The last section contains a summary and concluding remark.

## 2  $q$-Ddeformed Fermionic Oscillator Models

### 2.1  The FN-Oscillators

In this subsection, we present some of quantum statistical mechanical properties of the multi-dimensional $q$-deformed fermionic Newton oscillator algebra, called the FN-oscillators. The $U(d)$-covariant algebra generated by the FN-oscillators $c_i$ together with their corresponding creation operators $c_i^*$ is defined by the following deformed anti-commutation relations [51]:

$$c_i c_j^* + q\, c_j^* c_i = q^{\hat{N}} \delta_{ij}, \qquad i,j = 1,2,...,d,$$

$$c_i c_j + c_j c_i = 0, \qquad\qquad\qquad (1)$$

$$c_j \hat{N} = (\hat{N}+1)\, c_j,$$

where $\hat{N}$ is the total fermion number operator in $d$ dimensions, and $q \in R$, $q > 0$. This system has the total deformed fermion number operator for $d$-dimensional case $c_1^* c_1 + c_2^* c_2 + .... + c_d^* c_d = [\hat{N}]$, whose spectrum is given by

$$[N] = N\, q^{N-1}, \qquad\qquad (2)$$

where $N = N_1 + N_2 + ... + N_d$ and each $N_i$ can have eigenvalues of 0 and 1. From (1) and (2), the multi-dimensional undeformed fermionic oscillator algebra can be obtained in the limit $q = 1$. Also, one can check that the FN-oscillator algebra in (1) shows $U(d)$-symmetry. Under the linear transformation

$$c_i' = \sum_{j=1}^{d} T_{ij} c_j, \qquad\qquad (3)$$

the relations given in (1) are covariant. Here, the matrix $T \in U(d)$, and it satisfies the unitarity condition $T\overline{T} = 1$, where the matrix $\overline{T}$ is the adjoint matrix of *T*. This property justifies the name Newton. From the relations in (1), it also follows that the FN-oscillators fulfill the Pauli exclusion principle.

Since we shall concern the main differences among the one-dimensional *q*-deformed fermionic oscillators in the next sections, the one-dimensional case of the FN-oscillators will be of special interest. Historically, it is interesting to note that the fermionic algebra proposed by Chaichian and Kulish [3,4], namely,

$$cc^* + qc^*c = q^{\hat{N}},$$

$$c^2 = 0, \quad [N,c] = -c, \quad [N,c^*] = c^*, \tag{4}$$

arises as a particular one-dimensional case in the *U(d)*-covariant *d*-dimensional FN-oscillators in (1) and (2). As far as we know, the same type of the algebra has been independently investigated by Ng [5], who used a precondition on the spectrum of the fermionic number operator with $\{0,1\}$. Although, Hayashi [52] and Frappat *et al.* [53] proposed the multi-dimensional *q*-fermionic oscillator models, we refer the one-dimensional algebra in (4) as the CKN algebra. Therefore, the most commonly accepted one-dimensional definitions of the CKN algebra with $q \in R^+$, namely,

$$cc^* + qc^*c = q^N,$$

$$cc^* + q^{-1}c^*c = q^{-N}, \tag{5}$$

$$c^2 = (c^*)^2 = 0,$$

are strictly isomorphic to the usual one-dimensional fermionic oscillator algebra after rescaling the operators [11],

$$cc^* + c^*c = 1, \qquad c^2 = 0. \tag{6}$$

But, it should be emphasized that the FN-oscillators in (1) present a different generalized fermions with a spectrum given by deformed integer eigenvalues in (2), although its one-dimensional case is equivalent to the usual fermion oscillator in (6). In this context, instead of studying a one-dimensional system, the multi-dimensional FN-oscillators deserve a special interest in the framework of statistical mechanics.

For comparison, we now summarize some of the thermostatistical properties of a gas of the FN-oscillators defined in (1) and (2). In the grand canonical ensemble, the Hamiltonian of such a free FN-oscillators gas has the following form [54]:

$$\hat{H}_F = \sum_i (\varepsilon_i - \mu)\,\hat{N}_i, \tag{7}$$

where $\varepsilon_i$ is the kinetic energy of a particle in the state $i$, $\mu$ is the chemical potential, and $\hat{N}_i$ is the fermion number operator relative to $\varepsilon_i$. This Hamiltonian represents essentially a non-interacting system of the FN-oscillators since we do have neither a specific deformed anti-commutation relation between fermionic annihilation (or creation) operators, nor a quantum group symmetry structure in (1). Also, this Hamiltonian does implicitly incorporate deformation, since the occupation number depends on the deformation parameter $q$ by means of (1) and (2).

Following the procedure proposed in [9,29], from (1) and (2), we have the following relation including the thermal averages:

$$< c_i c_i^* > + q < c_i^* c_i > = < q^{\hat{N}} >, \tag{8}$$

which leads to the statistical distribution function $[f_i^q]$ for the FN-oscillators as [54]

$$[f_i^q] \equiv < c_i^* c_i > = \frac{q}{e^{\beta(\varepsilon_i - \mu)} + q}. \tag{9}$$

The usual Fermi-Dirac distribution can be recovered in the limit $q = 1$. Using (9), one can deduce the logarithm of the fermionic grand partition function of the system as

$$\ln Z_F = \sum_i \ln(1 + qze^{-\beta\varepsilon_i}), \tag{10}$$

which gives all of the thermodynamical functions in terms of the deformation parameter $q$. Following the standard procedure [55-57], we can determine the equation of state and the particle density for the FN-oscillators gas as follows [54]:

$$\frac{P}{kT} = \frac{1}{\lambda^3} f_{5/2}(q, z), \tag{11}$$

$$\frac{1}{\upsilon} = \frac{1}{\lambda^3} f_{3/2}(q, z), \tag{12}$$

where $\upsilon = V/N$ and $\lambda = \sqrt{(2\pi\hbar^2/mkT)}$ is the thermal wavelength. The generalized Fermi-Dirac functions $f_n(q, z)$ are defined as

$$f_n(q, z) = \sum_{l=1}^{\infty} (-1)^{l-1} \frac{(zq)^l}{l^n}. \tag{13}$$

These generalized functions reduce to the standard Fermi-Dirac functions $f_n(z)$ in the limit $q = 1$ [55-57]. The equation of state for the FN-oscillator system can also be derived from (11) and (12) as a virial expansion in the three-dimensional space:

$$\frac{P\upsilon}{kT} = 1 + \frac{1}{2^{5/2}}\left(\frac{\lambda^3}{\upsilon}\right) + ..., \tag{14}$$

where the virial coefficients are independent of the deformation parameter $q$ as in the case of an undeformed fermion gas. The internal energy and the entropy of the FN-oscillators gas can be found as [54]

$$\frac{U}{V} = \frac{3}{2}kT\frac{1}{\lambda^3}f_{5/2}(q,z), \tag{15}$$

$$\frac{S}{Nk} = \frac{5}{2}\frac{f_{5/2}(q,z)}{f_{3/2}(q,z)} - \ln z, \tag{16}$$

which give the same results for an undeformed fermion gas in the limit $q = 1$. In the low-temperature limit, the number density in (12) can be rewritten as

$$\frac{N}{V} = \frac{4\pi}{3}\left(\frac{2mkT}{h^2}\right)^{3/2}(\ln(qz))^{3/2}\left[1 + \frac{\pi^2}{8}(\ln(qz))^{-2} + ...\right]. \tag{17}$$

From this relation, we find the chemical potential $\mu$ in the zeroth approximation as $\mu \approx \varepsilon_F - kT\ln q$, where $\varepsilon_F$ has the same form as in the case of standard fermions [55-57]. This expression shows the $q$-dependence only at finite temperatures. In the next approximation, we obtain [54]

$$\mu \approx -kT\ln q + \varepsilon_F\left(1 - \frac{\pi^2}{12}\left(\frac{kT}{\varepsilon_F}\right)^2\right), \tag{18}$$

which is different from the one for an undeformed fermion gas for the cases $q > 1$ and $q < 1$. But, in the limit $q = 1$, it reduces to the standard chemical potential $\mu$ for an undeformed fermion gas. Also, the results in (9)-(18) are different from the results of [31-34].

## 2.2 The CKN-Oscillators

In this subsection, for comparison, we summarize some of quantum statistical mechanical properties of the CKN algebra defined in (5). The CKN algebra can be rewritten by the following relations [3-5]:

$$cc^* + q^{-1}c^*c = q^{-N},$$ (19)

$$[N,c] = -c, \qquad [N,c^*] = c^*,$$

which satisfies the Pauli exclusion principle in the sense that the Fock states is restricted to $n \in \{0,1\}$, and it also reduces to the usual fermion algebra in the limit $q=1$. Recently, Narayana Swamy investigated the thermostatistics of this kind of fermionic $q$-oscillators in a specific interval of the deformation parameter $0 \le q \le 1$ [32,33]. The number operator spectrum of this algebra was derived by the relation

$$\beta_n = \left(\frac{1-(-1)^n}{2}\right)q^{-n+1} = \{0,1,0,q^{-2},0,q^{-4},....\},$$ (20)

which shows a different spectrum from the one considered for the algebra in (5). The thermostatistics of the CKN algebra can be studied by obtaining the mean occupation number $n_i$ defined by the relation $\hat{n} = Tr(e^{-\beta H}c^*c)/Z$. From a similar Hamiltonian in (7), one can obtain the distribution function as [32,33]

$$n_i = \frac{q^{-1}}{e^{\beta(\varepsilon_i-\mu)} + q^{-1}}.$$ (21)

In the limit $q = 1$, $n_i$ will be the usual Fermi-Dirac distribution. We should note that it is strictly different from the one obtained in (9), since both the CKN- and the FN-oscillator models have originally different realizations of the $q$-deformed fermionic algebras in (19) and (1).

According to (21), in Fig. 1, the $q$-deformed distribution function $n$ for a gas of the CKN-oscillators is shown for low temperatures as a function of $x$ for several values of the deformation parameter $q$.

From a form of the logarithm of the fermionic grand partition function $\ln Z = \sum_i \ln(1 + q^{-1} z e^{-\beta \varepsilon_i})$, one can obtain many of the thermostatistical properties in the thermodynamical limit. For instance, the pressure and the mean density can alternatively be obtained from (9), (11)-(13) under the exchange of the deformation parameter $q$ by $q^{-1}$. From such relations, the virial expansion may be found as

$$\frac{P \upsilon}{kT} = 1 + \frac{1}{2^{5/2}} \left( \frac{\lambda^3}{\upsilon} \right) + \left( \frac{1}{8} - \frac{2}{3^{5/2}} \right) \left( \frac{\lambda^3}{\upsilon} \right)^2 + .... \qquad (22)$$

The virial coefficients are independent of $q$ [32,33]. The internal energy and the entropy can also be found from (15) and (16) under the exchange of the deformation parameter $q$ by $q^{-1}$. On the other hand, one can find the chemical potential $\mu$ of a gas of the CKN-oscillators from (17) and (18) under the exchange of the deformation parameter $q$ by $q^{-1}$. Therefore, one can deduce the chemical potential in the zeroth approximation $\mu = \varepsilon_F - kT \ln q^{-1}$, which shows that the $q$-deformation is just a finite temperature effect [32,33]. This result is also similar to the one obtained for the FN-oscillators in (18). In the last section, we shall compare these results with the results for other $q$-fermions under consideration.

## 2.3 The PVC-Oscillators

In this subsection, for comparison, we summarize some of quantum statistical mechanical properties of another generalized fermions proposed by Parthasarathy and Viswanathan [6] as a possible $q$-deformation of the fermionic oscillator algebra. In addition, Chaichian *et al.* [8] further investigated some of statistical properties of the same algebra. Therefore, following the idea of Narayana Swamy [34], we call the system as the PVC-algebra. The PVC algebra is defined by the following relations [6,25]:

$$cc^* + qc^*c = q^{-N},$$

$$[N,c] = -c, \quad [N,c^*] = c^*, \quad c^2 \neq 0, \tag{23}$$

which can not be reduced to (6) due to the relation $c^2 \neq 0$. This relation means that a given quantum state may be occupied by more than two $q$-fermions, which contrasts with the Pauli exclusion principle. However, the spectrum of the generalized number operator of the PVC algebra is given by the following fermionic basic number [6,25]:

$$[n]_q^F = \frac{q^{-n} - (-1)^n q^n}{q + q^{-1}}, \tag{24}$$

which implicitly implies that the Pauli principle can only be recovered in the limit $q = 1$. Therefore, for $q \neq 1$, the PVC algebra in (23) presents a different non-trivial fermionic $q$-oscillator algebra. Recently, Narayana Swamy established the $q$-fermionic

Jackson [58] Derivative (JD) stemming from the PVC algebra for any function $f(x)$ as follows [34]:

$$D_x f(x) = \frac{1}{x} \frac{f(q^{-1}x) - f(-qx)}{q + q^{-1}}, \qquad (25)$$

which does not reduce to the ordinary derivative in the limit $q = 1$. The JD is required for the $q$-calculus. As discussed in [34], in order to study the thermostatistics of a gas of the PVC-oscillators, one should employ the fermionic JD instead of the ordinary thermodynamic derivatives, so that the standard structure of thermodynamics remains unchanged.

The thermostatistics of the PVC-oscillators was investigated by Narayana Swamy in a specific interval $0 < q < 1$ [34]. Using the relation $\hat{n} = Tr(e^{-\beta H} c^* c)/Z$, and from a similar Hamiltonian in (7), the $q$-fermion distribution function is derived as [34]

$$n = n(\eta) = \frac{1}{2|\ln q|} \left| \ln \left( \frac{\left| e^{\eta} - q^{-1} \right|}{e^{\eta} + q} \right) \right|, \qquad (26)$$

where $\eta = \beta(\varepsilon - \mu)$, $q \neq 1$. The behavior of the function $n(\eta)$ was examined for different values of the deformation parameter $q$ in the interval $0 < q \leq 1$ [34]. The pressure for a gas of the PVC oscillators is also given by

$$\frac{P}{T} = \ln Z = \frac{(q + q^{-1})}{2 \ln q} \sum_i \ln\left(1 - z e^{-\beta \varepsilon_i}\right). \qquad (27)$$

This relation can be rewritten by means of the fermionic JD in (25) as [34]

$$\frac{P}{T} = \frac{1}{\lambda^3} h(5/2, z, q), \qquad (28)$$

where the generalized Fermi-Dirac function $h(n, z, q)$ is defined by

$$h(n, z, q) = \frac{1}{2 \ln q} \left( \sum_{k=1}^{\infty} (-1)^{k+1} \frac{(q \, z)^k}{k^{n+1}} - \sum_{k=1}^{\infty} \frac{(q^{-1} z)^k}{k^{n+1}} \right). \tag{29}$$

The behavior of the function $h(5/2, z, q)$ was examined for different values of the deformation parameter $q$ in the interval $0 < q < 1$ [34]. One may also obtain the following thermodynamical functions such as particle density, the internal energy, and the entropy for a gas of the PVC oscillators, respectively [34]:

$$\frac{N}{V} = \frac{1}{\lambda^3} h(3/2, z, q) \,, \tag{30}$$

$$U = \frac{3}{2 \lambda^3} V \, T \, h(5/2, z, q) \,, \tag{31}$$

$$\frac{S}{V} = \frac{g}{\lambda^3} \left( \frac{5}{2} h(5/2, z, q) - h(3/2, z, q) \right), \tag{32}$$

where $g$ is the multiplicity factor. In the last section, we shall compare these results with the results for other $q$-fermions reviewed in previous subsections.

## 2.4 The VPJC-Oscillators

In this subsection, we would like to invatigate the quantum algebraic properties of another generalized fermionic oscillator algebra introduced by Viswanathan *et al*. [7]. Also, Chaichian *et al*. [8] further investigated some of its statistical properties. Therefore, we call the algebra as the VPJC-oscillators model. This model is defined by the following relations [7]:

$$cc^* + qc^*c = 1,$$

$$[N, c] = -c, \qquad [N, c^*] = c^*, \tag{33}$$

where $N$ is the number operator and $q \in R^+$. This model reduces to the usual fermion oscillator algebra in the limit $q = 1$ without the Pauli exclusion princinple. Although, this model can be found by making a transformation [7,8,20] from the PVC-oscillators model in (23), it notably reveals different quantum algebraic properties as will be shown below. Following the procedure proposed by Narayana Swamy [32-34] on the PVC- and the CKN-oscillator models, we wish to give a detailed physical analysis on the VPJC-oscillators model, which constitutes not only less popular model but also, less studied one in the literature.

Now, we construct the Fock space representation of the VPJC-oscillators model. Let $|n\rangle$ be the Fock space basis, and the deformed fermion number operator $[\hat{N}] = c^*c$ satisfies the property $[\hat{N}]|n\rangle = g_n|n\rangle$, where $g_n$ represents the eigenvalue of the deformed number operator $[\hat{N}]$. The actions of the deformed fermion oscillators $c$ and $c^*$ on the states may be assumed as follows:

$$c|n\rangle = C_n|n-1\rangle, \qquad c^*|n\rangle = C_n'|n+1\rangle, \tag{34}$$

where $C_n$ and $C_n'$ are the constants to be determined later. From (33) and (34), we obtain the recurrence relation as

$$g_{n+1} = 1 - qg_n, \tag{35}$$

which gives the following relations upon the choice $g_0 = 0$ having a role for the fermionic ground state:

$$g_1 = 1,$$
$$g_2 = 1 - q,$$
$$g_3 = 1 - q + q^2, \qquad\qquad (36)$$
$$.......$$
$$g_n = 1 - q + q^2 - q^3 + .... + (-1)^{n-1} q^{n-1}.$$

Such a result can be redefined as

$$g_n = [n] = \frac{1 - (-1)^n q^n}{1 + q}, \qquad\qquad (37)$$

which is the $q$-fermionic basic number for the VPJC-oscillators model. From the above analysis, we also obtain the relations:

$$cc^* = [\hat{N} + 1] = 1 - q[\hat{N}], \qquad C_n = \sqrt{[n]}, \qquad C_n' = \sqrt{[n+1]}. \qquad (38)$$

The Fock states can be constructed by applying the fermionic creation operators ($c^*$) on the ground state successively as

$$|n\rangle = \frac{(c^*)^n}{\sqrt{[n]!}} |0\rangle, \qquad\qquad (39)$$

where $[n]! = [n][n-1][n-2].....[1]$. Moreover, the Pauli exlusion princible in the VPJC-oscillators model can be recovered only in the limit $q = 1$, since we have

$$\lim_{q \to 1} g_1 = 1, \qquad\qquad g_2 = 0, \qquad\qquad (40)$$

which implies $(c^*)^n = 0$ for $n > 1$ in (39). Hence, the Fock states for this model reduces to the states $|0\rangle, |1\rangle$. Therefore, we need not assume the condition $c^2 = 0$ for this model in contrast with the situation introduced in [8]. Also, we see that

$$\lim_{q \to 1}[n] = \frac{1}{2}\left\{1 - (-1)^n\right\}, \tag{41}$$

which takes the values 0,1 for even and odd $n$, respectively. This situation is different from the usual fermion algebra in (6). However, the PVC-oscillators model has the same behavior in the limit $q = 1$, which can be deduced from (24). On the other hand, as has been pointed out in [7], from (37) and (39), the positive norm condition on the state vectors can not be satisfied for even values of $n$ if $q > 1$. Therefore, we consider the interval $0 < q < 1$ for the deformation parameter $q$ in this model.

Although, similar arguments were carried out for both the PVC- and the CKN-oscillator models in [32-34], from all the above remarkable properties, we should emphasize that the VPJC-oscillators model shows a different generalized fermions, except in the limit $q = 1$. Moreover, the following bosonic $q$-oscillator algebra was first introduced by Arik and Coon [59]:

$$bb^* - qb^*b = 1, \qquad 0 < q < 1,$$

$$[b, \hat{N}] = b, \qquad [b^*, \hat{N}] = -b^*, \tag{42}$$

whose number operator spectrum was defined by the relation

$$[n] = \frac{1 - q^n}{1 - q}. \tag{43}$$

Hence, the above analysis also shows us how the VPJC-oscillators model differs from the Arik-Coon $q$-oscillator model.

Following the procedure proposed in [33,34] on the PVC- and the CKN-oscillator models, we now establish the Jackson derivative (JD) appropriate for the VPJC-oscillators algebra. One may have the holomorphic representation as

$$c \Leftrightarrow D_x, \qquad c^* \Leftrightarrow x. \qquad (44)$$

Hence, the VPJC-oscillators algebra in (33) can be rewritten as

$$D_x x + qx D_x = 1. \qquad (45)$$

To derive a solution for this equation, we first observe the following relation

$$x[\hat{N}+1] + qx[\hat{N}] = x, \qquad (46)$$

which can be expressed by means of (38). From (45) and (46), and using the property $[\hat{N}]x = x[\hat{N}+1]$, we deduce the following solution:

$$D_x = \frac{1}{x}\left[\hat{N}\right] = \frac{1}{x}\left(\frac{1-(-1)^{\hat{N}} q^{\hat{N}}}{1+q}\right). \qquad (47)$$

If we use the property $(-q)^{\hat{N}} f(x) = f(-qx)$ [28], this fermionic JD can also be expressed as

$$D_x f(x) = \frac{1}{x}\left(\frac{f(x) - f(-qx)}{1+q}\right), \qquad (48)$$

for any function $f(x)$. The fermionic JD does not reduce to the ordinary derivative in the limit $q = 1$. Recently, many of the mathematical properties of the fermionic derivative operator $D_x$ were studied by Schork [23]. But, above analysis is different from Schork's study [23], where the author just employed a replacement $q \equiv -\tilde{q}$ with $\tilde{q} > 0$ for the Arik-Coon $q$-oscillator model in (42) and (43). Moreover, the deformation parameter $q$ for the VPJC-oscillators model used in the present study is strictly positive. Such a notion was also pointed out in [20]. For the above analysis, we have used both quantum algebraic properties of the VPJC-oscillators model in (33)-(41) and the holomorphy relation as in (44)-(48). The fermionic JD in (48) plays a central role in the framework of mathematics and physics such that it is not only needed to define a consistent formulation of the fermionic $q$-calculus, but also it is required to study the thermostatistics of a gas of the VPJC-oscillators.

In order to compare the behaviors of statistical distribution functions of the above fermionic $q$-oscillator models, we now wish to consider the statistical distribution of the VPJC-oscillators models. To derive the mean occupation numbers of each energy level, we choose a similar form of the fermionic Hamiltonian as in (7). The mean value of the $q$-deformed occupation number $n_i$ is defined by [30]

$$[n_i] = \frac{1}{Z} Tr(e^{-\beta \hat{H}_F} [\hat{N}_i]) \equiv \frac{1}{Z} Tr(e^{-\beta \hat{H}_F} c_i^* c_i) \,, \qquad (49)$$

where $\beta = 1/kT$ , $k$ is the Boltzmann constant, $T$ is the temperature of the system, and $Z = Tr(e^{-\beta \hat{H}_F})$ is the partition function. After applying the cyclic property of the trace [9,10], and using the Fock space properties of the VPJC-oscillators algebra in (33) and (38), we obtain

$$\frac{[n_i]}{[n_i+1]} = e^{-\beta(\varepsilon_i - \mu)}. \qquad (50)$$

Since we have the following relations from (40) and (41):

$$\lim_{q \to 1}[n] = n, \qquad \lim_{q \to 1}[n+1] = 1 - n, \qquad n = 0,1, \qquad (51)$$

where we have dropped the subscript $i$ for the sake of simplicity. Thus, the expression in (50) reduces to the usual Fermi-Dirac distribution in the limit $q = 1$. From the definition of the $q$-fermionic basic number $[n]$ in (37) and using (50), we accordingly derive

$$n = n(\eta) = \frac{1}{|\ln q|} \left| \ln \left( \frac{|e^{\eta} - 1|}{e^{\eta} + q} \right) \right|, \qquad (52)$$

where $\eta = \beta(\varepsilon - \mu)$, $q \neq 1$. This equation provides the $q$-fermion distribution of the VPJC-oscillators model, which may also be called the $q$-deformed Fermi-Dirac statistical distribution function for a gas of the VPJC-oscillators.

Furthermore, the $q$-deformed distribution function $n(\eta)$ in (52) possesses some important properties as follows: (i) It satisfies the positivity condition to be the correct fermion distribution function. (ii) It is just a formal solution for the $q$-deformed fermion distribution in the interval $0 < q < 1$ due to the reasons mentioned after (41). However, one should consider the relations in (50) and (51) in order to find the usual Fermi-Dirac distribution as $n(\eta) = 1/(e^{\eta} + 1)$, which has a similar modified form for the case $q = 1$ for finite temperatures as shown in Fig. 1. (iii) It takes the standard step-functional form in the limit $T = 0$ for any values of $q$, just as in the cases of the FN-, CKN- and the PVC-oscillators. Hence, we conclude that the $q$-deformation of fermions is just a finite temperature effect even in the present VPJC-oscillators model. (iv) It is discontinuous at $\varepsilon = \mu$ and also, the peak of this function occurs at $\eta = 0$ for any values of $q$. (v) It vanishes at $\eta = -1.387$ for $q = 1/2$ and $\eta = -1.092$ for $q = 1/3$, respectively.

By means of the above considerations, in Fig. 2, the $q$-deformed distribution function $n(\eta)$ of the VPJC-oscillators is shown for finite temperatures as a function of $\eta = \beta(\varepsilon - \mu)$ for various values of the deformation parameters $q$. From the behavior of this function, we conclude that for $q \neq 1$, it is radically different from the behaviours of both the usual fermionic distribution and the three $q$-fermion models considered in previous subsections.

In the next section, we shall compare all the above properties presented by different fermionic $q$-oscillator models under consideration.

## 3. Discussion

We now wish to reveal similarities and differences among the four $q$-deformed fermion oscillator algebra models in the context of their quantum statistical mechanical properties.

We have discussed the fermionic quantum algebras of FN-oscillators, CKN-oscillators, PVC-oscillators and VPJC-oscillators. Particularly, we have focused on some general properties of the VPJC-oscillators model. Although it has some different and interesting properties, this model is not so popular in the literature comparing to the other models considered. We have determined the $q$-fermionic basic number which follows from this model. Our analysis is different from the studies of [7,8,20,23], since we pursued a careful and detailed analysis of the Fock space properties of the VPJC-oscillators model.

We may conjecture that this analysis may be used in a construction of the $N = 2$ SUSY algebra composed by both the VPJC-fermion and the Arik-Coon boson oscillators. The fact that when $q \neq 1$, the VPJC-oscillators model shows different generalized fermions without exclusion principle, i. e., the Fock states may be occupied by arbitrary number of quanta with $n = 0,1,2,3,....,\infty$, may be interpreted as follows: Such a behavior shows that $q$-deformed fermions behave like bosons for those values of $q$, and hence they lead to supersymmetrization in the sense that we can collect them together in the same Fock states of SUSY generators. This may also be interpreted as a unification of quantum oscillators, but the key point here is that $q$-deformation of fermions can lead to supersymmetry. The deformation parameter $q$ in this case can be thought of as a control parameter such that the original fermionic character of the system may be changed to a bosonic one in the specific interval $0 < q < 1$. On the other hand, the deformation parameter $q$ of the VPJC-oscillators model may also be interpreted as an interpolating object between fermionic and bosonic behaviors of the system. Such a behavior is also a character of anyonic systems with fractional statistics. Hence, we may speculate that the VPJC-oscillators model can be used for studying such exotic systems.

Among the four $q$-deformed fermionic oscillator models, the system of the FN-oscillators in (1) and (2) provides essentially an example of non-interacting multi-mode system of the $q$-deformed fermionic particles. The algebra of the FN-oscillators has notable properties: (i) It has *U(d)*-symmetry, and (ii) the deformation parameter $q$ has values in the interval $0 < q < \infty$. But, the one-dimensional case of the FN-oscillator algebra has a trivial representation, since it is isomorphic to the usual fermion algebra as is shown in (6). However, its multi-dimensional case reveals non-trivial and different generalized fermions. Since, the algebras of CKN-, PVC-, VPJC-oscillators in (19), (23), (33) have been investigated in a specific interval of the deformation parameter $0 < q < 1$, we compare the quantum statistical behavior of the FN-oscillators in the same interval. We should note that for all the algebras considered here, the quantum statistical mechanical properties are different from the results of the usual fermion oscillators except in the limit $q = 1$. When we compare with the entropy values of an undeformed fermion gas, the entropy of the FN-oscillator gas in (16) decreases for $q < 1$. A similar variation is seen in the pressure of the FN-oscillator gas such that in (11), the pressure is lower than the result of an undeformed fermion gas for $q < 1$ at the same fugacity. Both the pressure and the entropy values of the FN-oscillators gas in the interval $q < 1$ in (11) and (16) are lower than the values of the similar characteristics of the PVC-oscillators gas in (28) and (32).

Furthermore, the *q*-deformed Fermi-Dirac distribution function in (9) takes the standard step-functional form in the limit $T = 0$, just as in the cases of the CKN-, the PVC- and the VPJC-oscillators. However, for all the above generalized fermion systems, the results are different from the results of the standard Fermi-Dirac distribution function in the interval $0 < q < 1$. Thus, the *q*-deformation of fermions modifies the standard Fermi-Dirac distribution. Among the four deformed algebras, the Pauli exclusion principle satisfies only for the cases of FN- and CKN-oscillator algebras. The forms of the generalized Fermi-Dirac functions for the FN-, the CKN-, the PVC- and the VPJC-oscillator models are different, since their origins depend on the different definitions of $q$-fermions.

According to Fig. 1, the values of the $q$-deformed Fermi-Dirac distribution function $n$ of the CKN-oscillators model for the interval $0 < q < 1$ are larger than those of an undeformed fermion gas. Also, the value of this function for the case $q < 1$ increases when the deformation parameter is decreased. However, such a result is in contrast to the behavior of the $q$-deformed distribution function $[f_i^q]$ of the FN-oscillators model in (9). Strictly speaking, when the deformation parameter $q$ decreases, the values of the generalized function $[f_i^q]$ for the FN-oscillators decrease for the case $q < 1$ [54].

According to Fig. 2, the values of the $q$-deformed Fermi-Dirac distribution function $n(\eta)$ of the VPJC-oscillators model for the case $q < 1$ decrease when the deformation parameter $q$ is decreased. A similar variation has been observed in the behavior of the $q$-fermion distribution function of the PVC-oscillators model in [34]. However, both the PVC- and the VPJC-oscillator models employ different realizations of the one-dimensional fermionic $q$-oscillator algebras. Hence, the structures of the generalized distribution functions for these models have different behaviors for low temperatures as shown in (26) and (52).

Among the four deformed fermion oscillators, just for the PVC- and the VPJC-oscillator systems, one must use the fermionic JD in (25) and (48) instead of the usual thermodynamic derivatives to study the thermostatistics of a gas of such oscillators. Also, the FN- and CKN-oscillator algebras are not related to the fermionic basic numbers in contrast to the case of the PVC- and the VPJC-oscillator algebras. However, the situation is somewhat different used for the CKN algebra in [31], where the basic number similar to the $q$-bosons is assumed with the eigenvalues of $\{0,1\}$. Recently, the CKN algebra was also used to study both for obtaining its thermostatistical properties and for its possible relation to interpolating statistics in a specific interval $0 < q < 1$ [32,33]. Moreover, both the PVC- and the VPJC-oscillator algebras could be used to study fractional statistics due to their unusual Fock space representation properties, i. e. it is possible to occupy more than two $q$-fermions in a given quantum state. Also, the virial coefficients in (14) and (22) do not depend on the deformation parameter $q$. This situation contrasts with the results of [41-44], where a quantum group covariant two-parameter deformed fermion gas was considered, and the virial coefficients were expressed in terms of the two independent real deformation parameters. Also, the results in (14) and (22) are different from the results of [31], where the authors used the CKN algebra with a different fermion number operator spectrum.

## 4. Summary

In this work, we have studied four different $q$-deformed fermion oscillator algebras, whose underlying quantum statistics exhibit different characteristics than those of the usual fermions. As has been mentioned in [32], an outstanding problem is to address the question about what kind of algebra would be best candidate to study the interpolating statistics. According to the above analysis, it is possible to speculate that the FN-oscillators together with their bosonic version the so-called the $q$-deformed bosonic Newton oscillators [60] could be used for studying the interpolating statistics. Since the bosonic Newton oscillators (BN) showed interesting thermostatistical characteristics only in the region $q < 1$ [61], we believe that both the FN- and the BN-oscillators could be seen as possible candidates to approach the behavior of some complex systems such as anyons. Hence, we may infer that even in the two-spatial dimensions, the FN- and the BN-oscillator algebras may exhibit unusual statistical behaviors different from the undeformed theory. We hope that these problems will be addressed in the near future.

## Acknowledgments

The author wishes to thank the referee for useful suggestions. Also, the author thanks D. Irk for his help in preparing the two-dimensional plots.

**Figure Captions**

**Fig. 1.** The $q$-deformed Fermi-Dirac distribution of the CKN-oscillators as a function

of $x = \beta\varepsilon$ and $\xi = \beta\mu$ for various values of the deformation parameter $q$ for low temperatures.

**Fig. 2.** The $q$-deformed Fermi-Dirac distribution $n(\eta)$ of the VPJC-oscillators as a function of $\eta = \beta(\varepsilon - \mu)$ for various values of the deformation parameters $q$ for finite temperatures.

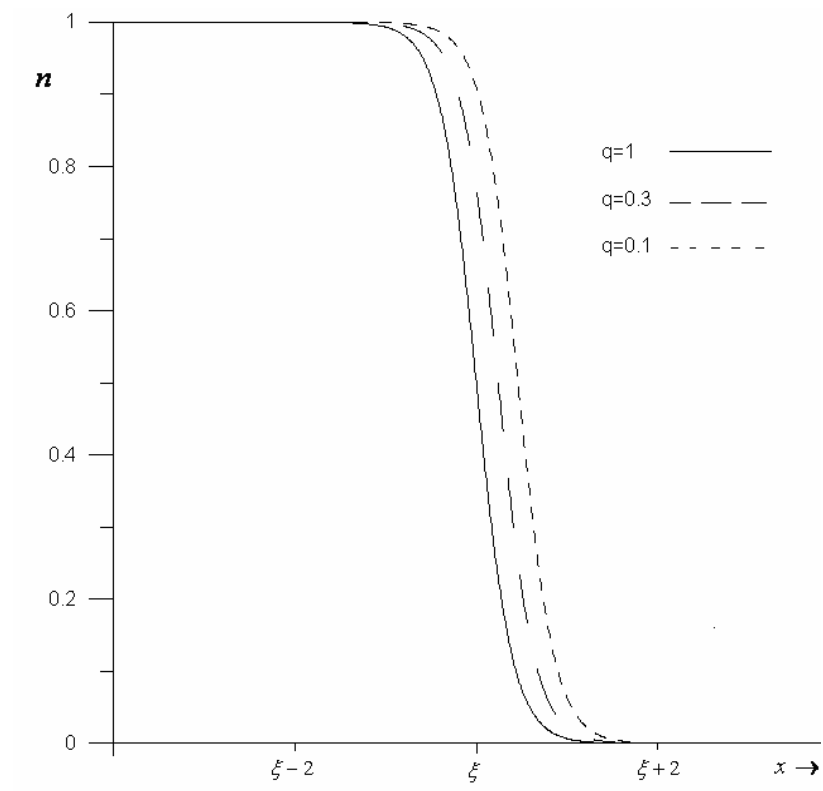

**Fig. 1**

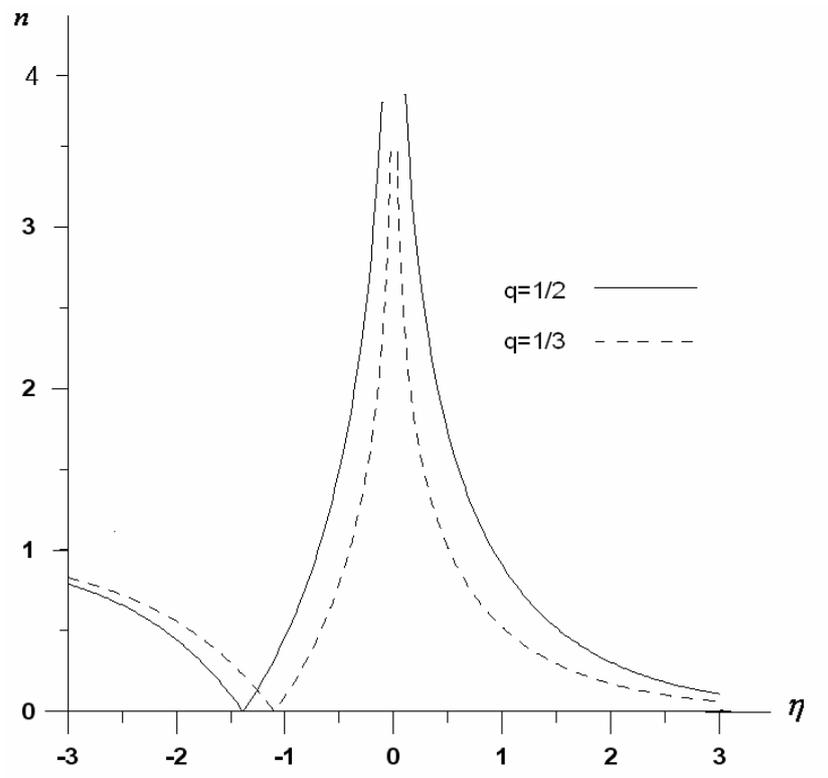

**Fig. 2**